\documentclass[aps,pre]{revtex4}
\usepackage{graphicx}
\usepackage{amssymb}
\usepackage{amsmath}
\usepackage{epsfig}

\begin{document}

\title{Comment on ``Long Time Evolution of Phase Oscillator Systems'' [Chaos 19,
023117 (2009)]}

\author{Edward Ott, Brian R. Hunt and Thomas M. Antonsen}

\affiliation
{University of Maryland, College Park, MD 20742}


\maketitle

Reference \cite{OA1} (henceforth referred to as $I$) considered
a general class of problems involving the evolution of large
systems of globally coupled phase oscillators. It was shown 
there that, in an appropriate sense, the solutions to these 
problems are
time asymptotically attracted toward a reduced manifold of system 
states (denoted $M$). This result has considerable utility
in the analysis of these systems, as has been amply demonstrated
in recent papers \cite{OA2}-\cite{Ghosh}. 
In this note, we show that the analysis of $I$
can be modified in a simple way that establishes significant
extensions of the range of validity of our previous result. 
In particular, we generalize $I$ in the following ways:
(1) attraction to $M$ is now shown for a very general class of
oscillator frequency distribution functions $g(\omega)$, 
and (2) a previous
restriction on the allowed class of initial conditions
is now substantially relaxed. In particular, with respect to point
(1), we now show that the result of $I$ (derived there for the
special case of Lorentzian $g(\omega)$) actually applies
for a very general distributions $g(\omega)$ described in the
third paragraph below.


To proceed, we recall the general class of problems described by Eqs.(1)
to (4) of $I$, where the time evolution of the quantity
$H(t)$ [appearing in Eq.(4) of $I$] is determined from the time evolution
of the order parameter $r(t)$ [given by Eq.(2) of $I$], and, as discussed
in $I$, this determination of $H$ from $r$ can arise in various ways
depending on the particular problem under consideration. The oscillator
distribution function $F(\theta,\omega,t)$ is then expressed 
[Eqs.(5) and (6) of $I$] in terms of a decomposition involving
functions $F_+(\theta,\omega,t)$ and $F_-(\theta,\omega,t)$,
where the analytic continuation of $F_+$ $(F_-)$ into 
$Im(\theta)>0$ $(Im(\theta) < 0)$ has no singularities and decays to 
zero as $Im(\theta) \rightarrow +\infty$ 
$(Im(\theta) \rightarrow -\infty)$. $F_+$ is further decomposed 
[Eq.(8) of $I$] into two parts, $F_+=\hat{F_+} + \hat{F'_+}$,
where $\hat{F'_+}$ lies on the reduced manifold $M$ and has dynamics 
described by Eq.(9) of $I$, while $\hat{F_+}$ satisfies 

\begin{equation}
\frac{\partial \hat{F_+}}{\partial t} + \frac{\partial}{\partial \theta}
\left\{ \left[
\omega + \frac{1}{2i} \left( H e^{-i\theta} - H^* e^{i \theta} \right)
\right]  \hat{F_+} \right\} = 0.
\label{eq:F_+_evol}
\end{equation}

\noindent
As argued in $I$, the time asymptotic evolution of the order parameter
will be completely described by dynamics on $M$ 
(i.e., by $\hat{F'_+}$) provided that

\begin{equation}
\lim_{t \rightarrow +\infty} 
\int_{-\infty}^{+\infty} \hat{F_+}(\theta,\omega,t) g(\omega) d\omega = 0,
\label{eq:F_+_cond}
\end{equation}

\noindent
where $g(\omega)$ is the oscillator frequency distribution function.
In $I$ it is shown that Eq.(\ref{eq:F_+_cond}) is satisfied for the case where
$g(\omega)$ is Lorentzian and the analytic continuation of the
initial condition, $\hat{F_+}(\theta,\omega,0)$, into 
$Im(\omega)<0$ has no singularities and approaches zero as 
$|\omega| \rightarrow \infty, Im(\omega) < 0$. Here we will show
that (\ref{eq:F_+_cond}) is satisfied much more generally.

We consider a general class of oscillator frequency distributions 
$g(\omega)$ that are analytic for real $\omega$ and can be analytically
continued into a strip $S$ defined by $0 \geq Im(\omega) > -\sigma$,
$\sigma > 0$, in which $g(\omega)$ has no singularities and decays to zero
as $Re(\omega) \rightarrow \pm \infty$ fast enough so that
$\int_{-\infty}^{+\infty} g(\omega_r + i \omega_i) d\omega_r$ is
well-defined. Some examples of $g(\omega)$ satisfying these conditions
are the Maxwellian, $g(\omega) \sim 
\exp \{-(\omega-\bar{\omega})^2/[2(\Delta \omega)^2]\}$, the Lorentzian, 
$g(\omega) \sim [(\omega-\bar{\omega})^2+(\Delta \omega)^2)]^{-1}$, 
and the sech distribution, 
$g(\omega) \sim \mbox{sech}[(\omega-\bar{\omega})/\Delta \omega]$.
In contrast, paper $I$ restricted consideration to Lorentzian 
$g(\omega)$. We further assume that the initial condition 
$\hat{F_+}(\theta,\omega,0)$ satisfies the same condition 
in $S$ as $g(\omega)$. Note that this condition on 
$\hat{F_+}(\theta,\omega,0)$ is considerably weaker than the condition
used in $I$ where it was required that $\hat{F_+}(\theta,\omega,0)$
be analytic everywhere in $Im(\omega)<0$ and 
$\hat{F_+}(\theta,\omega,0) \rightarrow 0$ for 
$|\omega| \rightarrow \infty, Im(\omega) < 0$.

In order to show (\ref{eq:F_+_cond}) for the above conditions
on $g$ and $\hat{F}_+$, we consider the solution
of Eq.(\ref{eq:F_+_evol}) for $\omega$ in the strip $S$.
Noting that $\omega$ only appears as a parameter in 
(\ref{eq:F_+_evol}) we can, for the moment, 
regard $\omega=\omega_r+i\omega_i$
$(\omega_i<0)$ as fixed. Furthermore, by means of the replacement, 
$\hat{F_+} \rightarrow \hat{F_+} e^{-i \omega_r t}$, we can,
without loss of generality, take $\omega_r=0$.
Thus Eq.(\ref{eq:F_+_evol})  takes the same form as Eqs.(17)
and (18) of $I$. Furthermore, it was shown [see 
Eqs.(19) to (31) of $I$ and the accompanying discussion] that
the solution of (18) and (19) of $I$ approaches zero at 
$t \rightarrow +\infty$. Thus, for any $\omega \in S$ and 
$Im(\omega) < 0$, we have that $|\hat{F_+}| \rightarrow 0$
as $t \rightarrow + \infty$. To complete our argument, we now note
that our conditions allow us to shift the integration path for the
integral in (\ref{eq:F_+_cond}) from the real $\omega$-axis to the 
line $\omega_r + i\omega_i$, $0>\omega_i>-\sigma$, 
$-\infty \leq \omega_r \leq +\infty$. Since at every point $\omega_r$
on this line $|g(\omega)|$ is bounded (because $g(\omega)$ has no
singularities in $S$) and $\hat{F_+} \rightarrow 0$ as 
$t \rightarrow \infty$, we conclude that Eq.(\ref{eq:F_+_cond})
is satisfied.

The above establishes a greatly expanded range of situation in which
Eq.(9) of $I$ can be used for discovering the time asymptotic dynamics 
of phase oscillator systems. We note, however, that, while (9) of
$I$ is substantially simpler than the original problem, it is still
nontrivial. Thus for the solution of (9) for the order
parameter dynamics it may still be useful to employ special choices
for $g(\omega)$ such as the Lorentzian (see Ref.\cite{OA2}).

Finally, we note that our requirement that $g(\omega)$ be analytic 
means that our proof of attraction to $M$ does not apply if 
$g(\omega)$ is a delta function. Thus attraction to $M$ necessitates
that $g(\omega)$ have a finite width. This is in line with the
intuition that relaxation to $M$ results from phase mixing
of oscillators in a population with heterogeneous frequencies.
The inapplicability of our result in the case where oscillators
all have the same frequency is also consistent with 
past results \cite{Pikovsky,Watanabe,Marvel2} which imply that,
when there is no frequency spread, the dynamics is not attracted 
to $M$.

\vspace{7mm}
This work was supported by ONR through grant N00014-07-1-0734.



\begin{thebibliography}{1}

\bibitem{OA1}
E. Ott and T.M. Antonsen, Chaos {\bf 19}, 023117 (2009).

\bibitem{OA2}
E. Ott and T.M. Antonsen, Chaos {\bf 18}, 037113 (2008).

\bibitem{Laing1}
C.R. Laing, Physica D {\bf 238},1569 (2009).

\bibitem{Lee}
W.S. Lee, E. Ott and T.M. Antonsen, Phys.Rev.Lett. {\bf 103} 044101 (2009).

\bibitem{Laing2}
C.R. Laing, Chaos {\bf 19}, 013113 (2009).

\bibitem{Abdulrehem}
M.M. Abdulrehem and E. Ott, Chaos {\bf 19}, 013129 (2009).

\bibitem{Marvel1}
S.A. Marvel and S. Strogatz, Chaos {\bf 19}, 013132 (2009).

\bibitem{Martens}
E.A. Martens, et. al., Phys.Rev.E {\bf 79}, 026204 (2009).

\bibitem{Pikovsky}
A. Pikovsky and M. Rosenblum, Phys.Rev.Lett. {\bf 101}, 264103 (2009).

\bibitem{Childs}
L.M. Childs and S.H. Strogatz, Chaos {\bf 18}, 043128 (2009).

\bibitem{So}
P. So, B.C. Cotton and E. Barreto, Chaos {\bf 18}, 037114 (2009).

\bibitem{Pazo}
D. Pazo and E. Montbrio, Phys.Rev.E {\bf 80}, 046215 (2009).

\bibitem{Ghosh}
A. Ghosh, D. Roy and V.K. Jirsa, Phys.Rev.E {\bf 80}, 041930 (2009).

\bibitem{Watanabe}
S. Watanabe and S.H. Strogatz, Phys.Rev.Lett. {\bf 70}, 2391 (1993);
Physica D {\bf 74}, 197 (1994).

\bibitem{Marvel2}
S.A. Marvel, R.E. Mirollo and S.H. Strogatz, Chaos {\bf 19}, 
043104 (2009).

\end{thebibliography}
\end{document}